\documentclass[aps,preprint,showpacs,superscriptaddress,groupedaddress]{revtex4}

\usepackage[T1]{fontenc}
\usepackage[latin1]{inputenc}

\usepackage{verbatim}
\usepackage{amsmath, multirow, amssymb, wasysym, gensymb}
\usepackage{dcolumn}   
\usepackage{bm}        
\usepackage{amssymb}   
\usepackage{amsmath}

\usepackage{graphicx}
\usepackage{amsfonts}
\usepackage{graphicx}
\usepackage{graphics}
\usepackage[ngerman, english]{babel}
\usepackage{float}
\usepackage{subfigure}
\usepackage{mathrsfs}

\usepackage{color}
\definecolor{carnelian}{rgb}{0.7, 0.11, 0.11}

\newcommand{\nue}{\nu_e}
\newcommand{\nuebar}{\bar{\nu}_e}

\begin{document}
\title{Big Bang Nucleosynthesis in the presence of sterile neutrinos with altered dispersion relations}
\author{Elke Aeikens}
\affiliation{University of Vienna, Faculty of Physics,
Boltzmanngasse 5, A-1090 Vienna, Austria}
\author{Heinrich P\"as}
\affiliation{Fakult\"at f\"ur Physik,
Technische Universit\"at Dortmund, 44221 Dortmund,
Germany}
\author{Sandip Pakvasa}
\affiliation{
Department of Physics \& Astronomy, University of Hawaii, 
Honolulu, HI 96822, USA}
\author{Thomas J. Weiler}
\affiliation{Department of Physics and Astronomy, Vanderbilt University, Nashville, TN 37235, USA}

\begin{abstract}
\noindent Big Bang Nucleosynthesis imposes stringent bounds on light sterile neutrinos mixing with the active flavors.
Here we discuss how altered dispersion relations can weaken such bounds and allow compatibility of new sterile neutrino
degrees of freedom with a successful generation of the light elements in the early Universe.
\end{abstract}
\maketitle

\noindent Additional gauge singlet "sterile" neutrinos with masses in the eV range have been discussed as solutions to various neutrino oscillation
anomalies, including the appearance of anti-electron neutrinos in a anti-muon neutrino beam at LSND \cite{Aguilar:2001ty}, the appearance of electron neutrinos in a muon neutrino beam at MiniBooNE 
\cite{Aguilar-Arevalo:2013pmq}, the deficit of electron neutrinos in 19 short baseline reactor experiments 
 \cite{Mention:2011rk} and the disappearance of electron neutrinos in the GALLEX and SAGE calibration runs with Cr-51 and Ar-37 
\cite{Giunti:2010zu},
(for a global analysis see \cite{Kopp:2013vaa}). Moreover, the absence of an upturn in the solar neutrino spectrum at low energies in several solar neutrino experiments has been advocated as evidence for a sterile neutrino with a mass in the mili- to centi-eV range \cite{deHolanda:2010am}.
Such sterile neutrinos are, however, somewhat in conflict with the non-observation of neutrino disappearance at other reactor or accelerator experiments~\cite{Abazajian:2012ys}.
Most recently, the IceCube experiment reported a bound on active-sterile neutrino oscillations at energies up to 20~TeV 
which is in conflict with the parameter space of these hints
\cite{TheIceCube:2016oqi}.
At present there are more than 20 experimental neutrino oscillation projects under development or consideration to clarify these puzzling anomalies (see e.g. \cite{lassere}).

On the other hand,
additional neutrino degrees of freedom contribute to the radiation content in the early Universe and thus lead to a faster expansion and consequently a higher temperature for the weak interaction freeze out. This  results in a larger neutron abundance, and consequently a larger Helium abundance. 
The excellent agreement of the predicted primordial abundances of light elements, in particular 
of Helium-4,  with observations are one of the major successes of big bang cosmology.
That this process doesn't spoil the successful prediction of the observed primordial element abundances poses a stringent bound on the number of neutrino species present in the early Universe (at $T\sim $~MeV)~(see e.g.\cite{Pospelov:2010hj,Beringer:1900zz,Steigman:2012ve}). 
Similar bounds have been derived from CMB data and the Large Scale Structure of the Universe 
(at the later time/temperature $T\sim$~eV).

It is thus interesting to explore mechanisms that suppress the sterile neutrino production in the early Universe.
One such possibility is provided by matter effects, which can be enhanced, e.g., by a lepton asymmetry 
reducing the active-sterile neutrino mixing, and consequently also the sterile neutrino production from neutrino oscillations \cite{Foot:1995bm}. 
In a manner similar to the matter effects due to a lepton asymmetry, Altered Dispersion Relations (ADR) can also 
result in a suppression of sterile-active mixing, and thus to a suppression of the population of sterile neutrinos before 
the freeze out of weak interactions; 
thus light sterile neutrinos may become compatible with Big Bang Nucleosynthesis (BBN). 
In this paper we analyze quantitatively this effect of ADRs, and find a favorable consequence for sterile neutrino model building.

A simple but sufficiently accurate estimate of the 
$^4$He abundance $Y(^4\text{He})$ in terms of the neutron-to-proton ratio
$n/p$ determines the temperature of Big-Bang Nucleosynthesis
$T_{BBN}$ \cite{Lit3}:
\begin{align}
&Y(^4\text{He})
=\frac{2\,n_n/n_p}{1+n_n/n_p},
\end{align}
where
\begin{align}
n_n/n_p\simeq \exp{[-\Delta m_{np}/T_{BBN}+(\mu_e-\mu_{\nu_e})/T_{BBN}]},
\end{align}
with $\Delta m_{np}$ being the mass difference between $n$ and $p$, and $\mu_e,\,\mu_{\nue}$ being the chemical potentials
of the electron $e^-$ and electron-neutrino $\nu_e$.
The reaction rates for the back-and-forth conversion of neutrons and protons are
\begin{equation}
\label{rates}
\nue+ n \leftrightarrow p+e^- \quad {\rm and}\quad \nuebar+p  \leftrightarrow n+e^+ \,.
\end{equation}
The rates for the two processes in (\ref{rates}) sum up to a rate
\begin{align} 
\Gamma_{BBN}=
2 \langle n_{e}\,\sigma(E_{e},p_{e})\,|v_{e}|\rangle\label{Bbn}
\end{align}
where $n_{e}$ is the electron or $\nue$ particle number density, 
$\sigma$ is  the reaction cross section for either process in \eqref{rates},
 and $|v_{e}|$ is the relative lepton speed,
and fall out of equilibrium for 
\cite{Lit3,Trodden:2004st}:
\begin{eqnarray}
\Gamma_{BBN}(T) \lesssim  H(T)\,,
\label{ab}
\end{eqnarray}
with the Hubble parameter given by
\begin{eqnarray}
\label{aa}
H(T,g_{eff}^{1/2})=\left(\frac{8\pi \rm G}{3}\rho_R \right)^{1/2}=\left(\frac{8\,\pi^3 }{90}\right)^{1/2}g_{\rm eff}^{1/2}\,\frac{T^2}{m_{Pl}}\label{5h}.
\end{eqnarray}
Here  G denotes Newton's constant, G$=m_{Pl}^{-2}$ with $m_{Pl}$ being the Planck mass, 
and $g_{\rm eff}$ is the effective number of degrees of freedom at temperature $T$. The temperature $T_{BBN}$ is determined when the reaction rate equals the Hubble parameter in eq.~\eqref{ab}.

The introduction of sterile neutrinos affects these processes in two ways. First, the 
effective number of degrees of freedom 
$g_{\rm eff}$ will be increased so that the Hubble parameter in eq. \eqref{aa} is increased as well. 
And second,
the sterile neutrinos will affect the number density and energy of active neutrinos
$\nu_a$ ($a = e, \mu, \tau$) and thus also the reaction rate $\Gamma_{BBN}$ in eq.~\eqref{Bbn}.

A change in the Hubble parameter would in turn alter the temperature
$T_{\rm BBN}$, and thus, via the neutron-to-proton ratio, alter the observed light element abundances.
This consequence can be avoided if sterile neutrinos $\nu_s$ would be kept out of equilibrium before the onset of Big Bang Nucleosynthesis,
suppressing their production. This either imposes stringent limits on the 
$\nu_s-{\nu_a}$ oscillation parameters $(\theta,\,\Delta m^2)$ or requires new physics, for example
a lepton asymmetry increasing the neutrino matter effect, which suppresses the effective mixing. 

In this paper we demonstrate that
an analogous suppression can be obtained by considering a third scenario in which the simple, ultra-relativistic dispersion relation 
\begin{align}
\label{seven}
E \simeq p + m^2/2p
\end{align}
 is altered (ADRs) by an additional term $A^{ADR}$
for sterile neutrinos. The most simple realization is to assume different propagation speeds for active and sterile neutrinos, with
\begin{equation}
A^{ADR}(T)=\pm \epsilon E = \pm 3.151\,\epsilon\, T
\end{equation}
added to Eq.~(\ref{seven}).
The sign of the term indicates Lorentz violating reduction or enhancement, respectively, as it arises in sterile neutrino shortcuts in extra dimensions or refraction, respectively (note the sign difference to \cite{Pas:2005rb}!).
We note that Coleman and Glashow have advocated the equivalence of species-specific limiting-velocities and (species-specific)Lorentz violation.
\cite{Coleman:1998ti}.

As a consequence,
the effective neutrino masses and mixing are
altered in a way similar to what happens when neutrinos propagate inside matter.
In fact, the new Lorentz violating ADR term $A^{ADR}$ and the matter potential in the early Universe \cite{Enqvist:1990ek} 
\begin{align}
A^{matter}(T)= \xi_a T^5= \sqrt{2}\,G_F\,n_\gamma\left(
-A_a\,T^2/M_W^2\right)\,, 
\label{5z}
\end{align}
add up to the total
potential 
\begin{align}
A(T)=A^{matter}(T) + A^{ADR}(T)= \xi_a T^5 \pm 3.151\,\epsilon\, T,
\label{Kc}
\end{align}
with $G_F$ being the  Fermi constant,  $M_W$ the mass of the $W$ boson and $n_\gamma=0.2404 \,T^3$ the photon density.
In contrast to the works \cite{Dolgov:2002wy,Enqvist:1990ek,Kirilova:1999xj}, 
we assume zero lepton asymmetry or that any lepton asymmetry is negligible and plays no role
and set the corresponding term, usually denoted by $L_a$, to zero.
The active-flavor ($a$) dependent numerical factors $A_a$ are determined by the plasma background at the time of BBN, consisting of neutrons and protons
(but negligible anti-baryons), equal numbers of electrons and positrons, neutrinos and antineutrinos, 
and photons which may be neglected as their coupling to neutrinos is so tiny, resulting in
$A_e \simeq 55.0$ \cite{Enqvist:1990ek}
and $A_{\mu,\tau} \simeq 15.3$ \cite{Notzold:1987ik,Foot:1995bm}
(compare also \cite{Hannestad:2012ky}).

As a consequence, the effective two-flavor Hamiltonian in vacuo
\begin{align} 
{\cal H} = \frac{\Delta m^2}{4E} \begin{pmatrix} -\cos 2 \theta & \sin 2 \theta\\ \sin 2 \theta & \cos 2 \theta
\end{pmatrix} 
\label{Ham}
\end{align}
describing the active-sterile neutrino oscillations via
\begin{align}
i \frac{d}{dt}\begin{pmatrix} \nu_a \\ \nu_s \end{pmatrix} = {\cal H} \begin{pmatrix} \nu_a \\ \nu_s \end{pmatrix}
\end{align}
is altered by an additional term, ${\cal H'}= {\cal H} + {\delta \cal H}$ with
\begin{eqnarray}
\delta {\cal H} = \begin{pmatrix}
A^{matter} & 0\\ 0& -A^{ADR}
\end{pmatrix}.
\label{deltaH}
\end{eqnarray}

The resulting effective two-flavor classical amplitude for oscillation becomes
\begin{eqnarray}
\sin^2 2\tilde{\theta}  =
\frac{ \sin^2(2\theta)}{\sin^2(2\theta)+ \cos^2(2\theta)
\left(\frac{2 E\cdot A(E)}{\Delta m^2\cos(2\theta)}-1\right)^2}\label{sintil}
\end{eqnarray}
where the potential energy $A(E)$ now includes both the matter effects of the hot dense early universe as well as 
an ADR.

The vacuum mixing angle $\theta$ by definition occurs at $A(E)=0$.
The resonance condition  
\begin{eqnarray}
\frac{2 E A(E)}{\Delta m^2 \cos(2\theta)}\bigg\vert_{res}=1
\label{rescondnew}
\end{eqnarray}
determines the resonance energy.
%
At the resonance energy $E_{\rm Res}$, the additional terms in (\ref{deltaH}) cancel the
difference of the diagonal entries in (\ref{Ham}).
Note that for energies much smaller than the resonance energy $E_{\rm Res}$, 
$ {\cal H}$ dominates ${\delta\cal H}$,
the change in the dispersion relation decouples, 
and the scenario discussed resembles the standard vacuum 
scenario with 
three active and one sterile neutrino.
Note further that for energies much higher than $E_{\rm Res}$, the diagonal term ${\delta\cal H}$ dominates ${\cal H}$,
the $\nu_s$ production is highly suppressed, 
which generates the desired effect in the early Universe, which will be explained in more detail below.
Further details of such models have been  worked out in 
\cite{Pas:2006si,Dent:2007rk,Hollenberg:2009ws,Hollenberg:2009bq,Marfatia:2011bw}.

The averaged 2-flavor active-sterile oscillation probability is given by
\begin{equation}
\langle P_{\nu_a \rightarrow\nu_s}\rangle =
\left\langle \sin^2 \left( \frac{\Delta { \cal H }}{2} t \right) \sin^2(2\tilde\theta) \right\rangle
=\frac{1}{2}\sin^2(2\tilde\theta)\,,
\label{Ka}
\end{equation}
where $\Delta {\cal H}$ is the difference of the ${\cal H}$ eigenvalues, equal to $\frac{\Delta \tilde{m}^2}{2E}$.

Following \cite{Foot:1995bm}, we assume that the initial sterile neutrino density vanishes. 
The rate of sterile neutrino production is then given by the interaction rate of active neutrinos multiplied with the 
averaged oscillation probability
$\langle P_{\nu_a \rightarrow \nu_s}\rangle$, i.e.
\begin{align} 
\Gamma_{\nu_s}=\langle P_{\nu_a \rightarrow \nu_s}\rangle\Gamma_{\nu_a}.
\label{Gammanus}
\end{align} 
The condition that sterile neutrinos do not come into equilibrium then becomes
\begin{equation}
\Gamma_{\nu_s}(T)  \lesssim H(T)
\label{5y},
\end{equation}
which has to hold for all temperatures $T>T_{\rm BBN}$.
The reaction rates of the active neutrinos are slightly flavor-dependent.
The $\Gamma_{\nu_a}$, with $a=e,\mu,\tau$,  are given by 
\cite{Enqvist:1991qj,Foot:1995bm}:
\begin{equation}
\Gamma_{\nu_a}=y_a G_F^2\,T^5 \,,
\label{55c}
\end{equation}
where $y_e=4.0$, $y_{\mu,\tau}=2.9$.

The reaction rates of active neutrinos freeze out when $\Gamma_{\nu_a}\lesssim H(T)$, i.e., at temperatures
$T_e=2.6\,\text{MeV}$ and $\;T_{\mu,\tau}=4.4\,$MeV, respectively~\cite{Enqvist:1991qj,Foot:1995bm}.

\begin{figure}
\vspace{-10mm}
\includegraphics[width=15cm]{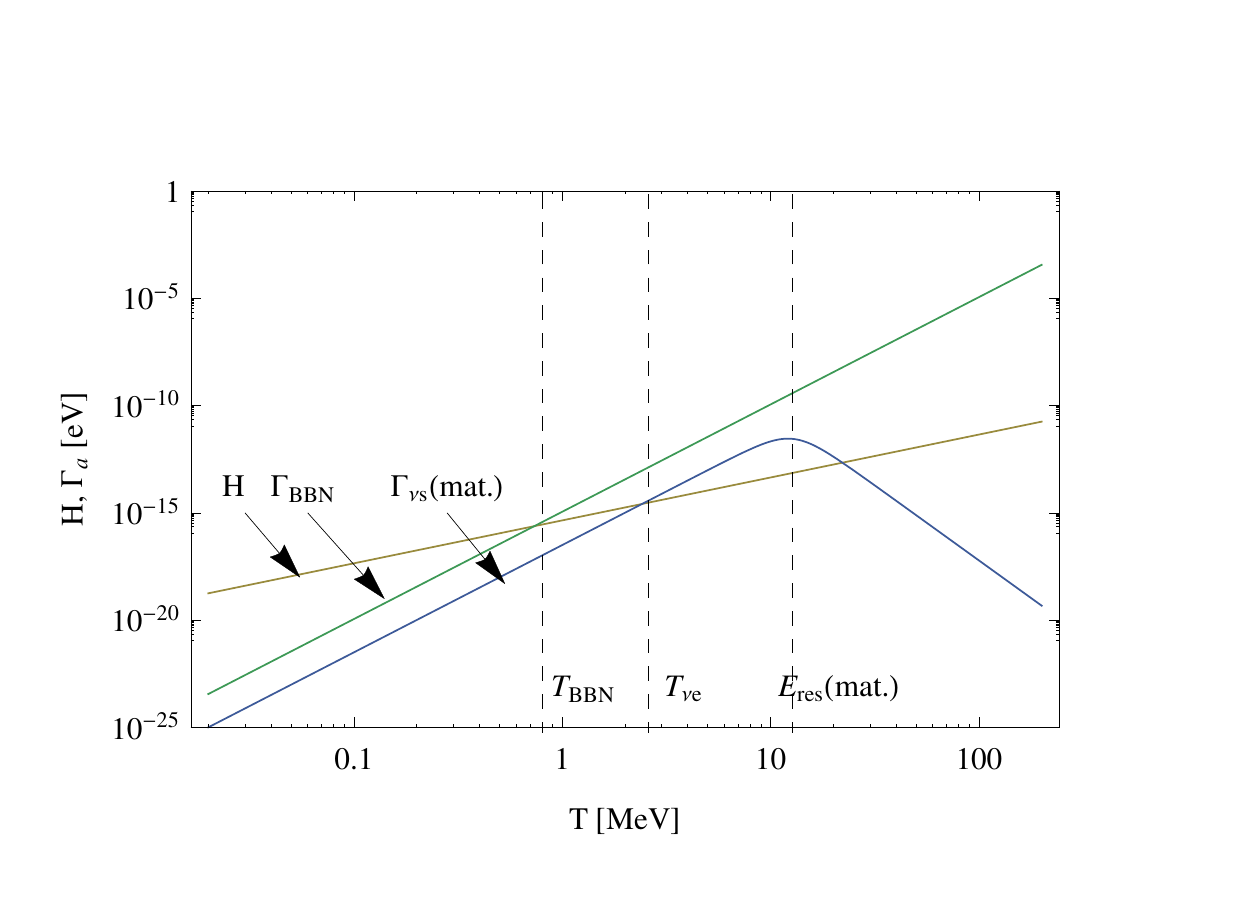}
\vspace{-15mm}
\caption{Reaction rates $\Gamma_{BBN}$, $\Gamma_{\nu_ s}$ as a function of temperature in comparison with the Hubble rate
$H$. BBN happens where $\Gamma_{BBN}$ falls below the Hubble rate. 
$\Gamma_{\nu_s}$ denotes the reaction rate
of sterile neutrinos with a matter potential for $\nu_e-\nu_s$ oscillations with $\sin^2\theta=0.03,\, \Delta m^2=0.93$ eV$^2$.}
\label{fig:A}
\vspace{-5mm}
\end{figure}

\begin{figure}
\vspace{-5mm}
\includegraphics[width=15cm]{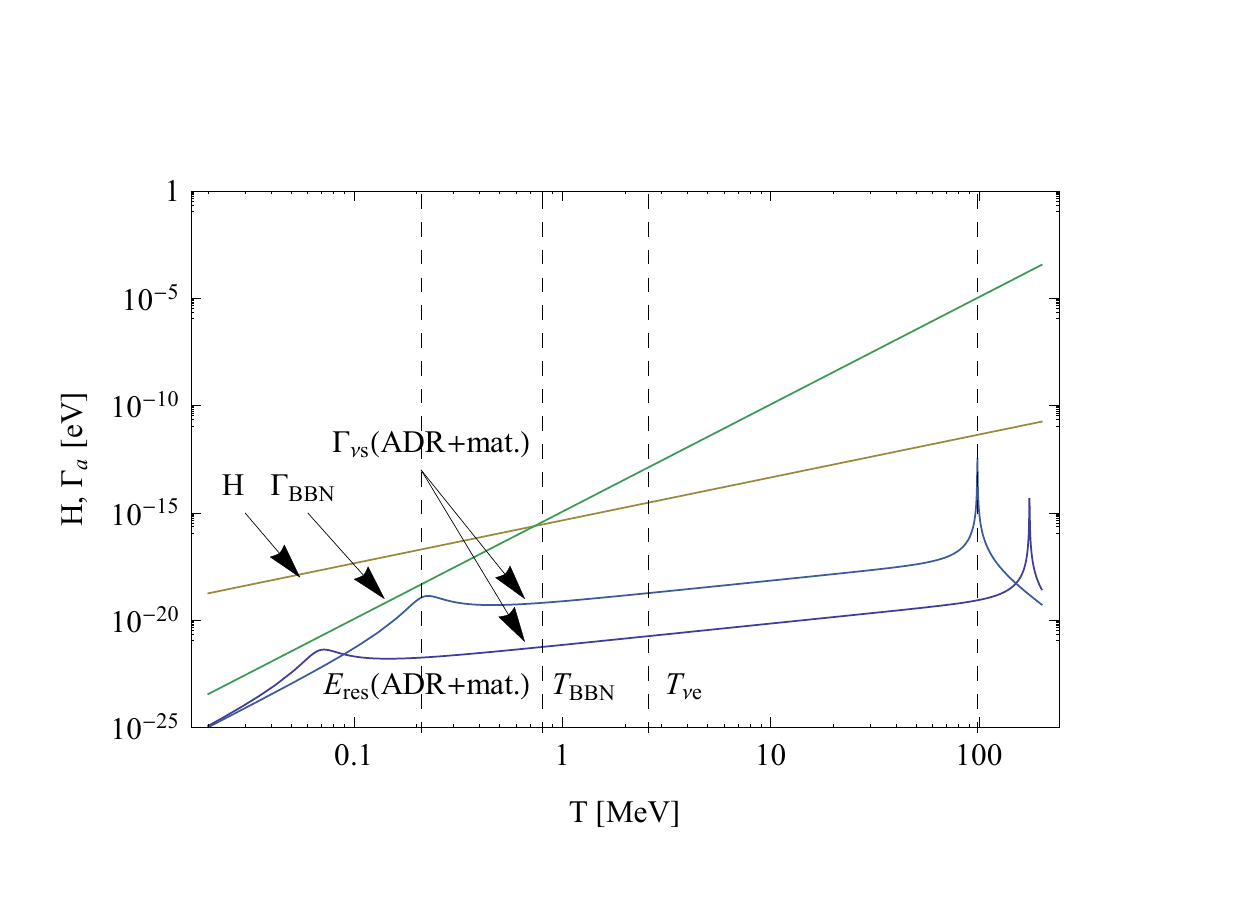}
\vspace{-15mm}
\caption{As above including both a matter potential and an altered dispersion relation due to shortcuts, $A=A^{ADR}+A^{matter}$, with 
$\epsilon=10^{-12}$ for the upper and $\epsilon=10^{-11}$ the lower blue (dark) curve, respectively.} 
\label{fig:C}
\vspace{-5mm}
\end{figure}

\begin{figure}
\vspace{-10mm}
\includegraphics[width=15cm]{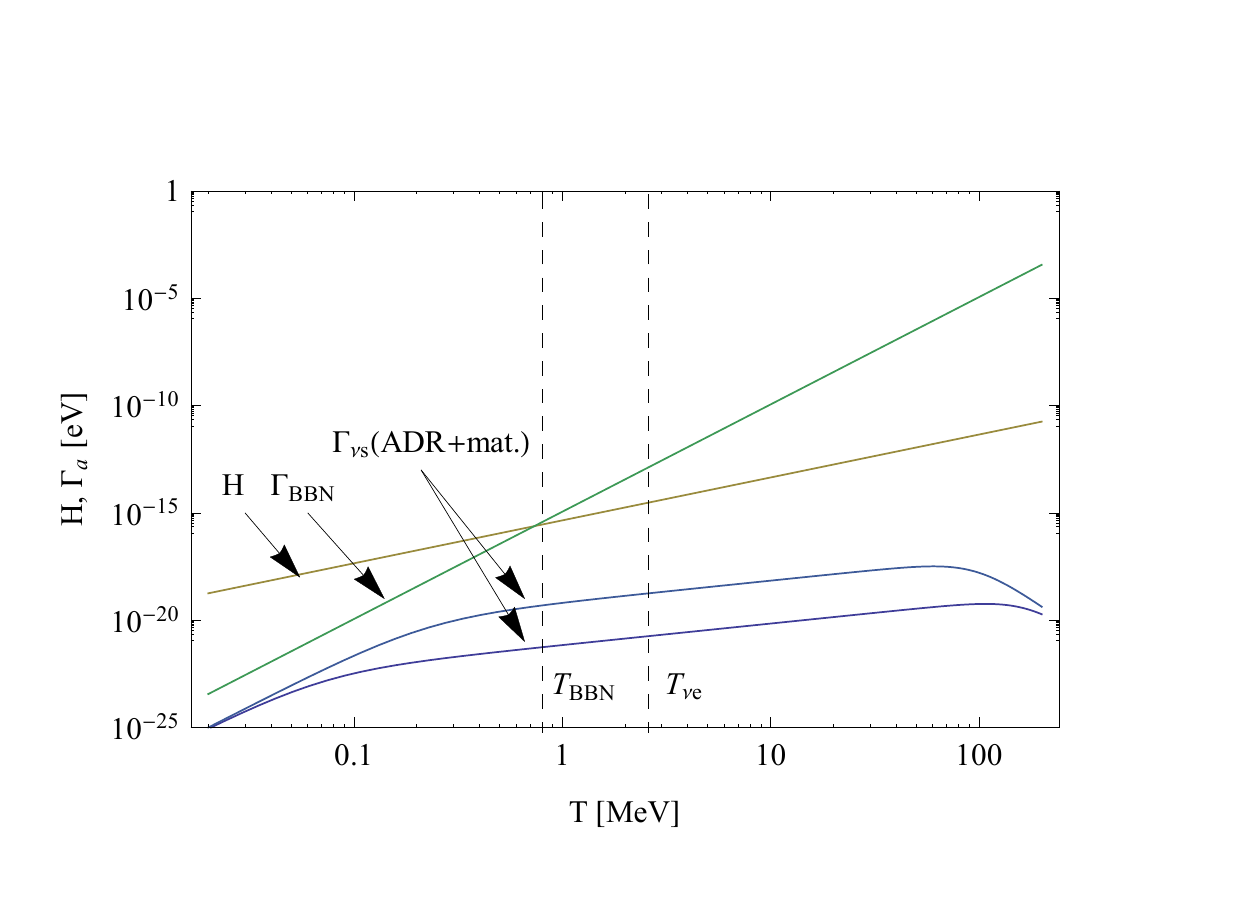}
\vspace{-15mm}
\caption{As above with a negative ADR potential $A=-A^{ADR}+A^{matter}$.}
\label{fig:E}
\end{figure}
 
Figs.~\ref{fig:A} to \ref{fig:E} show
the resulting 
sterile neutrino production rate $\Gamma_{\nu_s}$ (blue/dark) for oscillations of electron neutrinos as a function of temperature 
for an illustrative comparison
with the proton-neutron conversion rate $\Gamma_{BBN}$ (green/medium) and 
the Hubble parameter (yellow/light).
(Here, we have set $\sin^2\theta=0.03$ and $\Delta m^2=0.93$\,eV$^2$, 
corresponding to the best fit data of sterile neutrino oscillations according to~\cite{Kopp:2013vaa}.)
 
The crossing point of  $H$ with the BBN reaction rate $\Gamma_{BBN}$ 
defines the temperature $T_{\rm BBN} \,(\sim 0.8\, \rm MeV)$ at which BBN ceases to be effective, according to \eqref{ab}. 
The decoupling temperature of active neutrinos (in these figures the electron neutrinos) 
lies just above $T_{\rm BBN}$, at $T_{\nu_e}=2.6\,$MeV~\cite{Foot:1995bm}. 
 
Two peaks feature prominently in the sterile neutrino production rates in Fig.~\ref{fig:C}, where 
the matter and/or ADR potentials lead to amplified mixing and a change in the functional form which crucially affects the conditions for thermal equilibrium around BBN. While the first (low energy) peak is a real resonance peak, where the effective $\Delta \tilde{m}^2 \equiv 2 E \Delta {\cal H}$ vanishes, the
second (high energy) peak corresponds to the cancellation of matter potential and ADR only.
While the resonance condition eq.~(\ref{rescondnew}) is quadratic in energy and thus in principle has two solutions, the number of real positive solutions depends on the sign of $\Delta m^2$ and the ADR and matter potentials. 
We assume the sterile neutrino to be heavier than the active one.
Now the matter potential is negative, making active neutrinos effectively even lighter.
Thus the only possible resonance arises when a sterile neutrino shortcut cancels the mass difference between active and sterile
neutrinos.

Comparing the case for matter effects only (Fig.~\ref{fig:A}) with the case of matter effects plus an ADR potential
(Figs.~\ref{fig:C} and \ref{fig:E}), one can easily notice the following:

\begin{itemize}

\item In Fig.~\ref{fig:A} the resonance of the matter potential lies at
$E_{res}(\rm matter) = 13\,MeV$. $\Gamma_{\nu_s}$ is thus larger than the Hubble rate in the temperature interval
$T_{\nu_e}<T<22\,$MeV.
In this temperature interval, $\nu_s$-$\nu_e$ oscillations will populate $\nu_s$.
Thus, for this case, condition  \eqref{5y} is violated for $T> T_{\rm BBN}$.

\item 
In Figs.~\ref{fig:C} and \ref{fig:E}\, the 
oscillation probability $P_{\nu_e\rightarrow\nu_s}$ is shown  in the presence of an ADR, with $\epsilon=10^{-11}, 10^{-12}$, 
which yields an additional 
suppression factor of the active-sterile mixing. 
The cases shown correspond to a suppression (Fig.~2) or an enhancement (Fig.~3) of the sterile neutrino dispersion relation.
In Fig.~2, the total potential
$A=A^{matter}+A^{ADR}$ gives rise to the resonance energies
$E_{res}(\rm matter+ADR)\approx 0.2\,$MeV and 98.3\,MeV ($E_{res}\approx 0.07\,$MeV and 174.8\,MeV, respectively). 
The matter potential is less relevant here since
the ADR potential dominates. The maximum of the second resonance peak 
exceeds the Hubble rate only within the
small interval $T=(98.27-98.33)$\,MeV ($T=(174.79-174.80)$\,MeV) which corresponds to a time interval of 2\,ns and is less than the oscillation length of about 3\,ns (for $\epsilon=10^{-11}$). Therefore we expect the second resonance peak not to significantly populate sterile number densities. This condition poses another constraint in addition to eq.~\eqref{5y}, i.e. to suppress oscillations 
in the interval of the resonance peak $T=(T_1-T_2)$, the oscillation phase $\Phi[T]=\Delta \tilde{m}^2 t/4 E$ has to fulfill 
\begin{equation}
\Phi[T_2]-\Phi[T_1]<2\pi.
\label{Phicons}
\end{equation}
In Fig.~3, $\Gamma_{\nu_{s}}(T)<H(T)$ is fulfilled everywhere.

 \end{itemize}

To avoid an influence on the expansion rate of the Universe, the production rate of sterile neutrinos and antineutrinos 
has to be suppressed until the time of BBN $T_{BBN}$. 
Thus eq.  \eqref{5y} in combination with
eqs.~\eqref{5h}, \eqref{Ka}, \eqref{Gammanus}  and \eqref{55c}, 
evaluated at  $T_{BBN}$ imposes the constraint on the 
$\nu_s-\nu_a$ oscillation parameters:
\begin{align}
\sin^2(2\tilde\theta)\lesssim \frac{H}{\Gamma_{\nu_a}} =
	\left(\frac{8\pi^3}{90}\right)^{1/2}
	\frac{2 g_{\rm eff}^{1/2}}{y_a G_F^2 T^3\,m_{Pl}} \equiv \alpha_a.
	\label{zz}
\end{align}

Here, $g_{\rm eff}=10.75$, assuming that the number of degrees of freedom is unaffected by the sterile neutrinos, and
we find $\alpha_e\approx 3.20$ and $\alpha_{\mu,\tau}\approx 4.41$.
In addition, the condition \eqref{Phicons} has to be fulfilled for a positive ADR potential.

\begin{figure}
\includegraphics[width=13cm]{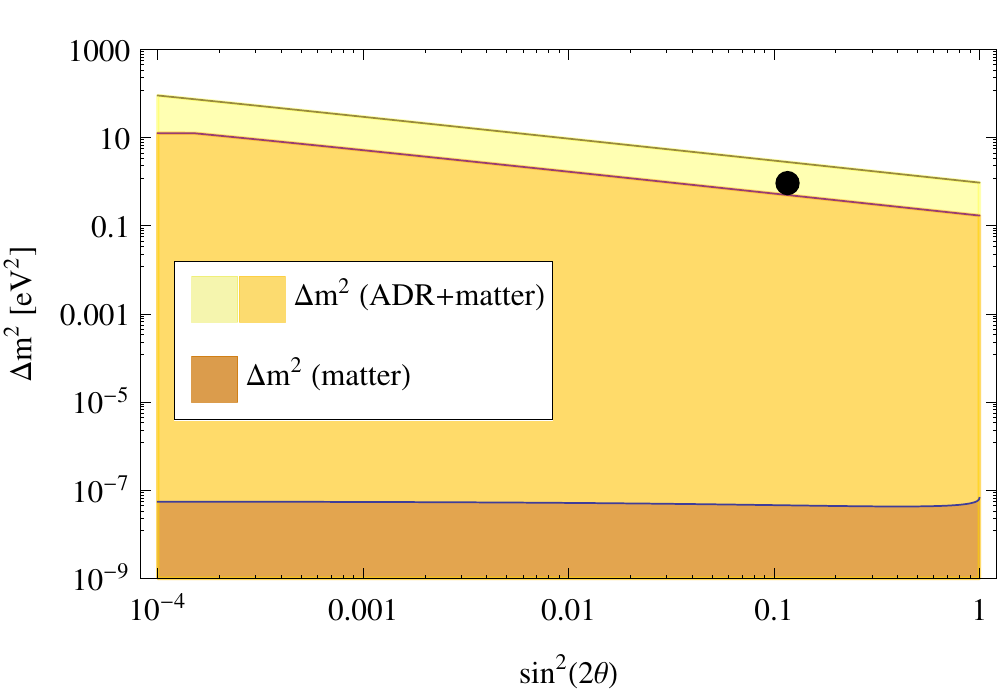}
\caption{Comparison of allowed regions (shaded areas) for successful BBN, in the two cases 
$\nu_s$-$\nu_e$ neutrino oscillations in matter, and oscillations 
in the presence of both matter and
an ADR potential $A=A^{matter} + \epsilon E$, for a shortcut parameter of $\epsilon=10^{-12}$(darker shaded) and  $\epsilon=10^{-11}$ 
(lighter shaded).}
\label{fig:Bildx1}
\includegraphics[width=13cm]{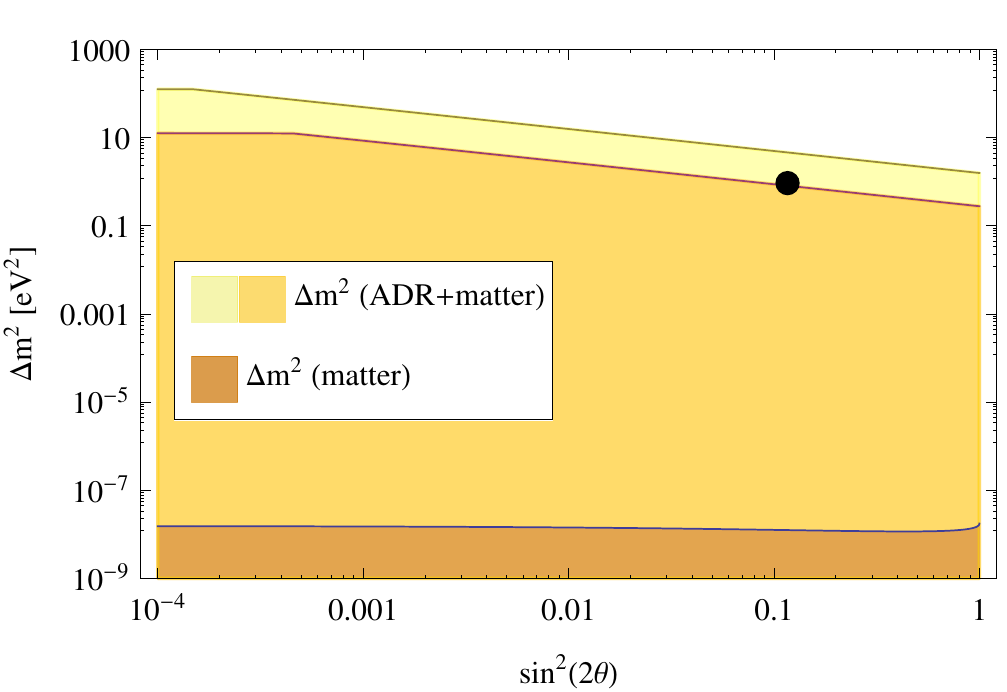}
\caption{As above, but for $\nu_s$-$\nu_{\mu/\tau}$ neutrino oscillations. }
\label{fig:Bildx4}
\end{figure}

\begin{figure}[h!]
\includegraphics[width=13cm]{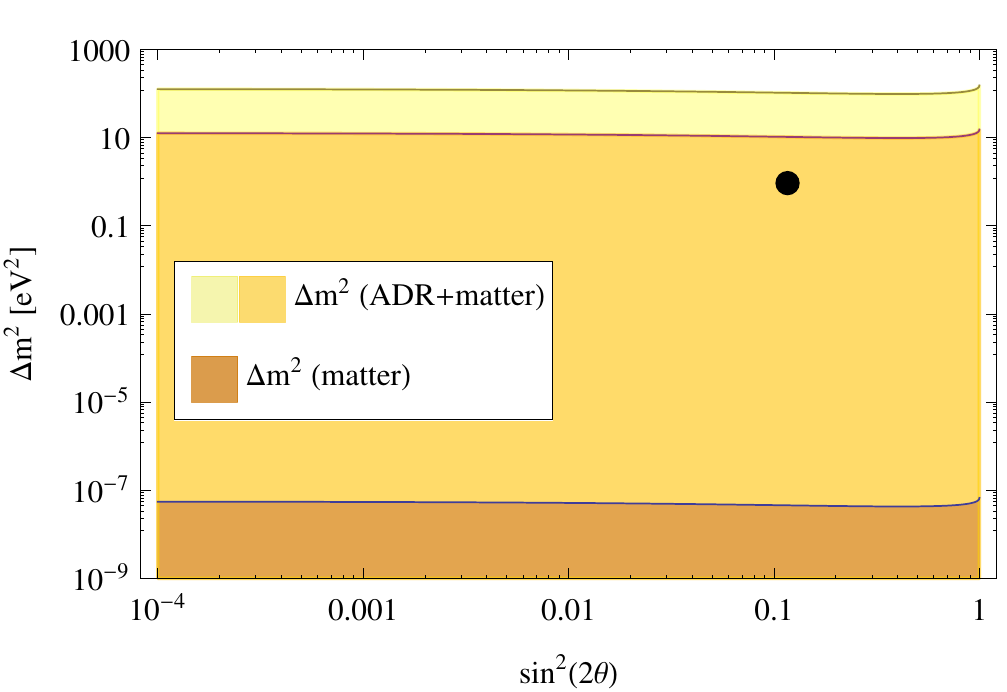}
\caption{As above but for $\nu_s$-$\nu_e$ neutrino oscillations with negative
ADR potential, $A=A^{matter}-\epsilon E$}
\label{fig:Bildx2}
\includegraphics[width=13cm]{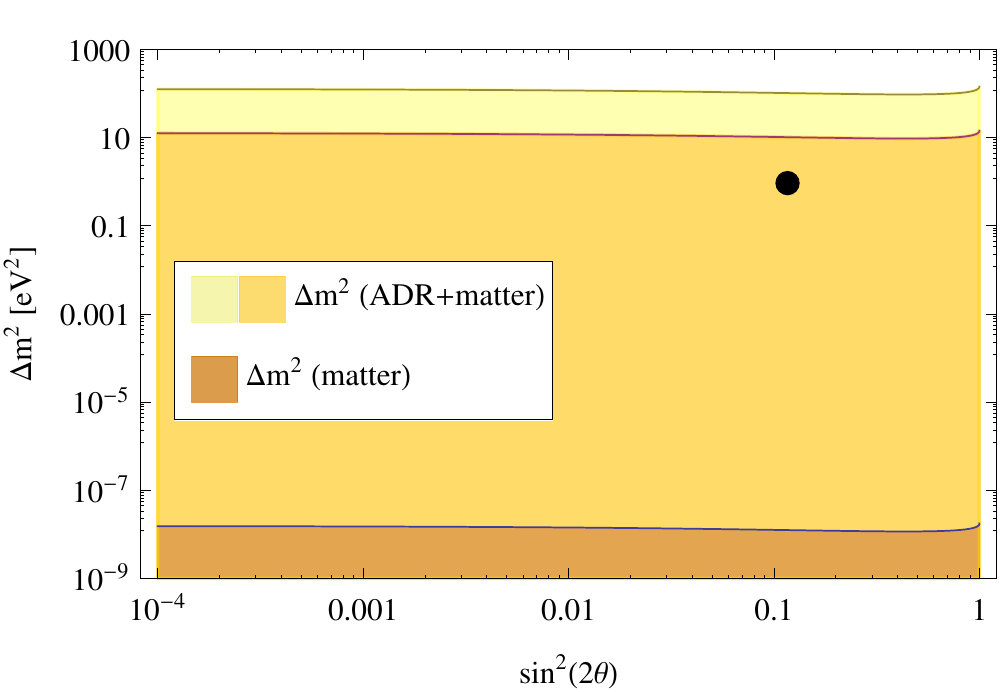}
\caption{As above, but for $\nu_s$-$\nu_{\mu/\tau}$ neutrino oscillations. }
\label{fig:Bildx3}
\end{figure}

Via eqns. \eqref{Ka} and \eqref{zz}, this implies the following relation between the sterile-active
$\Delta m^2$ and the vacuum mixing angle $\sin(2\theta)$
\footnote{Note that one of the earliest bounds placed on $\Delta m^2$ and mixing for a light light sterile neutrino from BBN is 
\cite{Langacker:1989sv}.}

\begin{eqnarray}
\Delta m^2\leq\frac{2 A(T_{BBN})E(T_{BBN})}
{\cos(2\theta)\pm\sin(2\theta)
\sqrt{(1-\alpha_a)/\alpha_a}}.\label{zzz}
\end{eqnarray}
The constraints from \eqref{zzz} and \eqref{Phicons} are displayed in Figs.~\ref{fig:Bildx1} and \ref{fig:Bildx4} for $\nu_e$-$\nu_s$ 
and  $\nu_{\mu,\tau}$-$\nu_s$ oscillations, respectively, with $\epsilon=10^{-12}$ and $10^{-11}$ and in Figs.~\ref{fig:Bildx2} and \ref{fig:Bildx3} 
for a  negative ADR potential.
The shaded areas depict the parameter space where
sterile neutrinos are not populated and BBN can proceed successfully. 
The parameter regions above the shaded areas are excluded. 
In each figure, cases for matter effects only (dark shaded areas), and for a combination of matter effects 
and an ADR potential (light shaded areas) are shown.
In the latter case the allowed region
is larger,
($\Delta m^2\sin^2(2\theta)\lesssim 0.71\,(0.28)\,$eV$^2$\, for $\nu_e\,(\nu_{\mu/\tau})$ with $\epsilon=10^{-12}$, 
and $\Delta m^2\sin^2(2\theta)\lesssim 0.95\,(1.58)\,$eV$^2$ for $\nu_e\,(\nu_{\mu/\tau})$ with $\epsilon=10^{-11}$, respectively), since the ADR potential 
$A^{ADR}=3.151~  \epsilon~ T_{BBN} \sim 10^{-6}$\,eV alone is sufficient to suppress the oscillation amplitude
$\sin^2(2\tilde\theta)$. In the case of pure matter effects
($A^{matter}[T_{BBN}]\sim 10^{-14}$\,eV) the allowed region is constrained to 
$\Delta m^2 \sin^2(2\theta) \lesssim 6.5\,(1.7)\cdot 10^{-8}\,$eV$^2$ for $\nu_e\,(\nu_{\mu/\tau})$.

In comparison to the case for $\nue$-$\nu_s$ oscillations  shown in Fig.~\ref{fig:Bildx1},
the bounds on $\nu_{\mu,\tau}-\nu_s$ oscillations in Fig.~\ref{fig:Bildx4}
are slightly less stringent due to a larger interaction rate
($\Gamma_{\mu,\tau}/\Gamma_{e}=1.38$).
Finally the allowed parameter regions become even larger when a negative ADR potential (postive refraction of sterile neutrinos) is assumed, as can be seen in Figs.~\ref{fig:Bildx2}, \ref{fig:Bildx3}.

The black dots correspond to the best fit data 
of the global neutrino oscillation data analysis including short and long-baseline accelerator, reactor, and radioactive source experiments, as well as atmospheric and solar neutrinos in a 3+1 scenario
\cite{Kopp:2013vaa}, with $\sin^2\theta=0.03$ and $\Delta m^2=0.93$\,eV$^2$. As can be seen, the global best fit value can be
made compatible with successful BBN in the early Universe in all cases by assuming a large enough shortcut parameter $\epsilon$.

Finally, one has to discuss the relevance of the first (low energy) resonance peak in  Fig.~\ref{fig:C} for the relic neutrino background.
In contrast to the high energy peak this peak is a real resonance that can lead to MSW transitions once the
adiabaticity condition
\begin{equation}
 \gamma[T_{peak}]=\frac{\tau_{sys}}{\tau_{int}}=\frac{1}{\omega_{eff}}\frac{d\tilde\theta}{dt}\ll 1\label{a}
\end{equation}
is fulfilled. The expression for the adiabaticity parameter (compare  \cite{Kuo:1989qe} with \eqref{sintil})
\begin{equation}
\gamma
 =\frac{8 \,(3.151\,T)^2\cdot H[T]}{(\Delta m^2)^2}\frac{\sin(2\theta)\,(A^{ADR}[T]+3A^{matter}[T])}{\left[\sin^2(2\theta)+\left(\cos(2\theta)-\frac{A\,2(3.151\,T)}{\Delta m^2}\right)^2\right]^{3/2}}
 \end{equation}
 results in
$ \gamma(\nu_e-\nu_s)\approx10^{-10}$, 
 $\gamma(\nu_{\mu/\tau}-\nu_s)\approx10^{-11}$, thus leading to virtually complete MSW conversion with an MSW probability
 $P_{\nu_a \rightarrow\nu_s}\approx 0.97$ which does not significantly depend on
$\sin\theta$ and $\Delta m^2$. Thus, while this transition is at temperatures low enough not to affect BBN 
or the total amount of neutrinos in the Universe $N_\nu$ anymore, 
depending on which and how many flavors mix with the sterile neutrino, the scenario predicts a
partly or virtually completely sterile relic neutrino background, making the difficult endeavor 
to detect the neutrino background 
\cite{relic-nu-exp}
even more challenging.

In summary, we have demonstrated
that ADR potentials yield a suppression of active-sterile neutrino mixing at high energies that has the potential to significantly
enhance the parameter space allowed for sterile neutrinos. 
Thus ADR scenarios such as shortcuts in extra dimensions 
\cite{Pas:2005rb} with an ADR parameter $\epsilon=10^{-12},\,10^{-11}$ allow an alternative to the case of large lepton asymmetries 
in order to make sterile neutrinos compatible with BBN.

\section*{Acknowledgments}
\noindent AE is supported by the FWF Austrian Science Fund under the Doctoral
Program W1252-N27 Particles and Interactions.
The research of TJW is supported in part by DoE grant DE-SC0011981.
HP is supported by DGF Grant No. PA 803/10-1 and
thanks the Alexander von Humboldt Foundation, Vanderbilt University and the University of Hawai'i at Manoa for additional financial support.


\end{document}